\documentclass[12pt]{iopart}

\usepackage{graphicx}

\begin{document}

\title{Entangling nanomechanical oscillators in a ring cavity by feeding squeezed light}
\author{Sumei Huang and G. S. Agarwal}
\address{Department of Physics, Oklahoma State University,
Stillwater, OK - 74078, USA} \eads{\mailto{sumei@okstate.edu},
\mailto{agirish@okstate.edu}}
\date{\today}

\begin{abstract}
A scheme is presented for entangling two separated nanomechanical
oscillators by injecting broad band squeezed vacuum light and laser
light into the ring cavity. We work in the resolved sideband regime.
We find that in order to obtain the maximum entanglement of the two
oscillators, the squeezing parameter of the input light should be
about 1. We report significant entanglement over a very wide range
of power levels of the pump and temperatures of the environment.
\end{abstract}

\vspace*{1in} \noindent{Contents}
\begin{enumerate}
\item[1.] Introduction\hspace{\fill}2
\item[2.] Model\hspace{\fill}3
\item[3.] Radiation pressure and quantum fluctuations\hspace{\fill}6
\item[4.] Entanglement of the two movable mirrors\hspace{\fill}8
\item[5.] Conclusions\hspace{\fill}12
\end{enumerate}
Acknowledgments\hspace{\fill}12\\ References\hspace{\fill}12
\maketitle
 \section {Introduction} It
is well known that entanglement is a key resource for quantum
information processing \cite{Nielsen}. One now has fairly good
understanding of how to produce entanglement among microscopic
entities. In recent times there has been considerable interest in
studying entanglement in mesoscopic and even microscopic systems
\cite{Vedral,Deb,Sen,Bose,Chou}. Nanomechanical oscillators are
beginning to be important candidates for the study of quantum
mechanical features at mesoscopic scales. In fact the possibility of
entangling two nanomechanical oscillators has been investigated from
many different angles: such as entangling two mirrors in a ring
cavity \cite{Mancini}, entangling two mirrors of two independent
optical cavities driven by a pair of entangled light beams
\cite{Zhang}, entangling two mirrors by using a double-cavity set up
by driving with squeezed light \cite{Pinard}, entangling two mirrors
of a linear cavity driven by a classical laser field \cite{Jpa},
entangling two mirrors in a ring cavity by using a phase-sensitive
feedback loop \cite{Ribichini}, entangling two dielectric membranes
suspended inside a cavity \cite{Plenio}, and entangling two
oscillators by entanglement swapping \cite{Pirandola,Vacanti}. Other
proposals do not use cavity configurations but coupling to Cooper
pair boxes \cite{Sbose}. Here we report a conceptually simple method
to produce entanglement between two mirrors. Our proposal enables us
to trace the physical origin of entanglement.

  In this paper, we propose a scheme for entangling two movable
mirrors of a ring cavity by feeding broad band squeezed vacuum light
along with the laser light. The two movable mirrors are entangled
based on their interaction with the cavity field. The achieved
entanglement of the two movable mirrors depends on the degree of
squeezing of the input light, the laser power, and the temperature
of the movable mirrors. The feeding of the squeezed light has been
considered to produce squeezing of a nanomechanical mirror
\cite{Jaehne,Sumei}. Further Pinard \textit{et al}. \cite{Pinard}
have considered entanglement of two mirrors in a double cavity
configuration which is fed by squeezed light - one part of the
cavity is fed by light squeezed in amplitude quadrature and the
other is fed by light squeezed in phase quadrature. In contrast we
consider a single mode ring cavity driven by a single component
amplitude squeezed light. In our scheme the entanglement can be managed
by an externally  controllable field which is the squeezed light.

The paper is organized as follows. In section 2 we introduce the
model, give the quantum Langevin equations, and obtain the
steady-state mean values. In section 3 we derive the stability
conditions, calculate the mean square fluctuations in the relative
momentum and the total displacement of the movable mirrors. In
section 4 we analyze how the entanglement of the movable mirrors can
be modified by the squeezing parameter, the laser power, and the
temperature of the environment. The parameters chosen in the paper
are from a recent experiment on optomechanical normal mode splitting
\cite{Aspelmeyer}.

 Before we present
our calculations, we present a key idea behind our work. For a
bipartite system, a sufficient criterion for entanglement is that
the sum of continuous variables satisfies the inequality \cite{Duan}
\begin{equation}\label{1}
\langle(\Delta (q_{1}+q_{2}))^2\rangle+\langle(\Delta
(p_{1}-p_{2}))^2\rangle<2,
\end{equation}
where $q_{j}$ and $p_{j}$ ($j=1,2$) are the position and momentum
operators for two particles, respectively. They obey the commutation
relation $[q_{j},p_{k}]=i\delta_{jk}$ ($j,k=1,2$).

Mancini \textit{et al}. \cite{Mancini} have derived another
sufficient condition for bipartite entanglement, which requires the
product of continuous variables satisfies the inequality
\begin{equation}\label{2}
\langle(\Delta (q_{1}+q_{2}))^2\rangle \langle(\Delta
(p_{1}-p_{2}))^2\rangle<1.
\end{equation}
In this paper, we will use equation (\ref{2}) to show the
entanglement between the two oscillating mirrors. Thus if we have a
situation where the interaction occurs only via the relative
coordinates $q_{1}-q_{2}$,$p_{1}-p_{2}$, then we can hold
$\langle(\Delta (q_{1}+q_{2}))^2\rangle$ at its value, says
$\simeq1$, before interaction and if the interaction can make
$\langle(\Delta (p_{1}-p_{2}))^2\rangle<1$, then the inequality (2)
would imply that the mirrors 1 and 2 are entangled. In the next
section we discuss how this can be achieved by using a single mode
ring cavity.

\section{Model}
\begin{figure}[!h]
\begin{center}
\scalebox{0.8}{\includegraphics{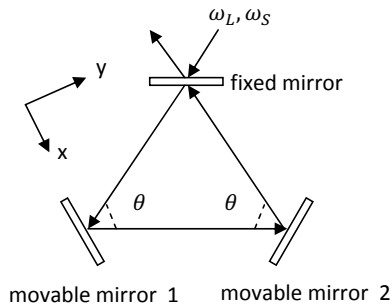}} \caption{\label{Fig1} Sketch
of the studied system. A laser with frequency $\omega_{L}$ and a
squeezed vacuum light with frequency $\omega_{S}$ enter the ring
cavity through the partially transmitting mirror.}
\end{center}
\end{figure}

The system under study, sketched in figure ~\ref{Fig1}, is a ring
cavity with one fixed partially transmitting mirror and two movable
perfectly reflecting mirrors, driven by a laser with frequency
$\omega_{L}$. As the photons in the cavity with length $L$ bounce
off the movable mirrors, they will exert a radiation pressure force
on the surfaces of the movable mirrors proportional to the
instantaneous photon number in the cavity. The motion of the movable
mirrors induced by the radiation pressure changes the cavity's
length, and alters the intensity of the cavity field, which in turn
modifies the radiation pressure force itself. Thus the interaction
of the cavity field with the movable mirrors through the radiation
pressure is a nonlinear effect. In addition, each mirror undergoes
quantum Brownian motion due to its coupling to its own independent
environment at the same low temperature $T$. The two movable mirrors
are identical with the same effective mass $m$, mechanical frequency
$\omega_{m}$ and momentum decay rate $\gamma_{m}$, and each mirror
is modeled as a quantum mechanical harmonic oscillator. We further
assume that the cavity is fed with squeezed light at frequency
$\omega_{S}$.

In the adiabatic limit, the cavity field is a single mode with
frequency $\omega_{c}$ \cite{Law}, and we can neglect the
retardation effect \cite{Aguirregabiria}, neglect the photon
creation in the cavity with moving boundaries due to the Casimir
effect \cite{Calucci}, and neglect the Doppler effect \cite{Karrai},
thus the radiation pressure force does not depend on the velocity of
the movable mirrors. Assuming the collisions of the photons on the
surfaces of the movable mirrors are elastic, the momentum
transferred to the mirrors per photon is $\hbar k_{y}-(-\hbar
k_{y})=2\hbar k_{y}$ (see figure ~\ref{Fig1} for the direction of
$y$), where $k_{y}=k\cos(\theta/2)$, $k$ is the wave vector of the
cavity field with $k=\omega_{c}/c$, and $\theta$ is the angle
between the incident light and the reflected light at the surfaces
of the movable mirrors. During the cavity round-trip time $t=2L/c$,
there are $n_{c}\cos(\theta/2)$ photons hitting on the surfaces of
the movable mirrors, so the radiation pressure force is
$F=n_{c}\cos(\theta/2)\times2\hbar k_{y}/t=n_{c}\hbar
\frac{\omega_{c}}{L}\cos^2(\theta/2)$. In a reference frame rotating
at the laser frequency, the Hamiltonian that describes the system
can be written as
\begin{equation}\label{3}
\eqalign{H=\hbar(\omega_{c}-\omega_{L})n_{c}+\hbar g
n_{c}\cos^2(\theta/2)
(Q_{1}-Q_{2})+\frac{\hbar\omega_{m}}{2}(Q_{1}^2+P_{1}^2)\\\hspace*{0.35in}+\frac{\hbar\omega_{m}}{2}(Q_{2}^2+P_{2}^2)+i\hbar\varepsilon(c^{\dag}-c),}
\end{equation}
 we have defined dimensionless position and momentum operators for the
oscillators $Q_{j}=\sqrt{\frac{m\omega_{m}}{\hbar}}q_{j}$ and
$P_{j}=\sqrt{\frac{1}{m\hbar\omega_{m}}}p_{j}$ ($j$=1,2) with
$[Q_{j},P_{k}]=i\delta_{jk}$ ($j,k=1,2$). Further in equation
(\ref{3}), $n_{c}=c^{\dag}c$ is the number of the photons inside the
cavity, $c$ and $c^{\dag}$ are the annihilation and creation
operators for the cavity field with $[c,c^\dag]=1$. The parameter
$g=\frac{\omega_{c}}{L}\sqrt{\frac{\hbar}{m\omega_{m}}}$ is the
optomechanical coupling constant between the cavity field and the
movable mirrors in units of s$^{-1}$. The different signs in front
of $Q_{1}$ and $Q_{2}$ are because the radiation pressure forces
exerted on the two mirrors are opposite. The parameter $\varepsilon$
is the coupling strength of the laser to the cavity field, which is
related to the input laser power $\wp$ by
$\varepsilon=\sqrt{\frac{2\kappa \wp}{\hbar\omega_{L}}}$, where
$\kappa$ is the photon decay rate by leaking out of the cavity.

In the system, the cavity field is damped by photon losses via the
cavity output mirror at the rate $\kappa$, and the movable mirrors
are damped due to momentum losses at the same rate $\gamma_{m}$.
Meanwhile, there are two kinds of noises affecting on the system.
One is the input squeezed vacuum noise operator $c_{in}$ with
frequency $\omega_S=\omega_L+\omega_m$. It has zero mean value, and
nonzero time-domain correlation functions \cite{Gardiner}
\begin{equation}\label{4}
\begin{array}{lcl}
\langle\delta c_{in}^{\dag}(t)\delta
c_{in}(t^{\prime})\rangle=N\delta(t-t^{\prime}),\vspace*{.1in}\\
\langle\delta c_{in}(t)\delta
c_{in}^{\dag}(t^{\prime})\rangle=(N+1)\delta(t-t^{\prime}),\vspace*{.1in}\\
\langle\delta c_{in}(t)\delta
c_{in}(t^{\prime})\rangle=Me^{-i\omega_{m}(t+t^{\prime})}\delta(t-t^{\prime}),\vspace*{.1in}\\
\langle\delta c_{in}^{\dag}(t)\delta
c_{in}^{\dag}(t^{\prime})\rangle=M^{\ast}e^{i\omega_{m}(t+t^{\prime})}\delta(t-t^{\prime}).
\end{array}
\end{equation}
where $N=\sinh^2(r)$, $M=\sinh(r)\cosh(r)e^{i\varphi}$, $r$ and
$\varphi$ are respectively the squeezing parameter and phase of the
squeezed vacuum light. For simplicity, we choose $\varphi=0$. The
other is quantum Brownian noises $\xi_{1}$ and $\xi_{2}$, which are
from the coupling of the movable mirrors to their own environment.
They are mutually independent with zero mean values and have the
following correlation functions at temperature $T$
\cite{Giovannetti}:
\begin{equation}\label{5}
\langle \xi_{j} (t)\xi_{k}
(t^{'})\rangle=\frac{\delta_{jk}}{2\pi}\frac{
\gamma_{m}}{\omega_{m}}\int\omega e^{-i\omega
(t-t^{'})}\left[1+\coth(\frac{\hbar \omega}{2k_{B}T})\right]d\omega,
\end{equation}
where $k_B$ is the Boltzmann constant and $T$ is the temperature of
the mirrors' environment, $j,k=1,2$.

The dynamics of the cavity field interacting with the movable
mirrors can be derived by the Heisenberg equations of motion and
taking into account the effect of damping and noises, which gives
the quantum Langevin equations
\begin{equation}\label{6}
\begin{array}{lcl}
\dot{Q_{1}}=\omega_{m}P_{1},\vspace*{.1in}\\
\dot{Q_{2}}=\omega_{m}P_{2},\vspace*{.1in}\\
\dot{P_{1}}=-g n_{c}\cos^2(\theta/2)-\omega_{m}Q_{1}-\gamma_{m}P_{1}+\xi_{1},\vspace*{.1in}\\
\dot{P_{2}}=g n_{c}\cos^2(\theta/2)-\omega_{m}Q_{2}-\gamma_{m}P_{2}+\xi_{2},\vspace*{.1in}\\
\dot{c}=-i[\omega_{c}-\omega_{L}+g\cos^2(\theta/2)
(Q_{1}-Q_{2})]c+\varepsilon-\kappa c+\sqrt{2
\kappa}c_{in},\vspace*{.1in}\\
\dot{c}^{\dag}=i[\omega_{c}-\omega_{L}+g\cos^2(\theta/2)
(Q_{1}-Q_{2})]c^{\dag}+\varepsilon-\kappa
c^{\dag}+\sqrt{2\kappa}c_{in}^{\dag}.
\end{array}
\end{equation}

From the second term of equation (\ref{3}), we can see only the
relative motion of the two movable mirrors is coupled to the cavity
field via radiation pressure. On introducing the relative distance
and the relative momentum of the movable mirrors by
$Q_{-}=Q_{1}-Q_{2}$ and $P_{-}=P_{1}-P_{2}$, we find that equation
(\ref{6}) reduces to
\begin{equation}\label{7}
\begin{array}{lcl}
\dot{Q}_{-}=\omega_{m}P_{-},\vspace*{.1in}\\
\dot{P}_{-}=-2g n_{c}\cos^2(\theta/2)-\omega_{m}Q_{-}-\gamma_{m}P_{-}+\xi_{1}-\xi_{2},\vspace*{.1in}\\
\dot{c}=-i[\omega_{c}-\omega_{L}+g\cos^2(\theta/2)
Q_{-}]c+\varepsilon-\kappa c+\sqrt{2
\kappa}c_{in},\vspace*{.1in}\\
\dot{c}^{\dag}=i[\omega_{c}-\omega_{L}+g\cos^2(\theta/2)
Q_{-}]c^{\dag}+\varepsilon-\kappa
c^{\dag}+\sqrt{2\kappa}c_{in}^{\dag}.
\end{array}
\end{equation}
We would use standard methods of quantum optics \cite{Walls} which have been adopted for discussions of quantum noise of nanomechanical mirrors \cite{Jpa,Giovannetti,DVitali,Paternostro1,Paternostro2},
setting all the time derivatives in equation (\ref{7}) to zero, and
solving it, we obtain the steady-state mean values
\begin{equation}\label{8}
P^{s}_{-}=0,\hspace{.02in}Q^{s}_{-}=-\frac{2g
|c^{s}|^{2}\cos^2(\theta/2)}{\omega_{m}},\hspace{.02in}c^{s}=\frac{\varepsilon}{\kappa+i\Delta},
\end{equation}
where
\begin{equation}\label{9}
\Delta=\omega_{c}-\omega_{L}+g Q^{s}_{-}\cos^2(\theta/2)
\end{equation}
is the effective cavity detuning, depending on $Q^{s}_{-}$. The
$Q^{s}_{-}$ denotes the new equilibrium relative distance between
the movable mirrors. Further $c^{s}$ represents the complex
amplitude of the cavity field in the steady state. For a given input
laser power, $Q^{s}_{-}$ and $c^{s}$ can take three distinct values,
respectively. Therefore, the system displays an optical
multistability \cite{Dorsel,Meystre,Marquardt}, which is a nonlinear
effect induced by the radiation pressure.

\section{Radiation pressure and quantum fluctuations}
To investigate entanglement of the two movable mirrors, we have to
calculate the fluctuations in the relative momentum of the movable
mirrors. This fluctuations can be calculated analytically by using
the linearization approach of quantum optics \cite{Walls}, provided
that the nonlinear effect between the cavity field and the movable
mirrors is weak. We write each operator of the system as the sum of
its steady-state mean value and a small fluctuation with zero mean
value,
\begin{equation}\label{10}
Q_{-}=Q^{s}_{-}+\delta Q_{-},\hspace*{.1in}P_{-}=P^{s}_{-}+\delta
P_{-},\hspace*{.1in}c=c^{s}+\delta c.
\end{equation}
Inserting equation (\ref{10}) into equation (\ref{7}), then assuming
the cavity field has a very large amplitude $c^{s}$ with
$|c^{s}|\gg1$, one can obtain a set of linear quantum Langevin
equations for the fluctuation operators,
\begin{equation}\label{11}
\begin{array}{lcl}
\delta\dot{Q}_{-}=\omega_{m}\delta P_{-},\vspace*{.1in}\\
\delta\dot{P}_{-}=-2g\cos^2(\theta/2)(c^{s\ast}\delta c+c^{s}\delta
c^{\dag})-\omega_{m}\delta Q_{-}-\gamma_{m}\delta P_{-}+\xi_{1}-\xi_{2},\vspace{.1in}\\
\delta\dot{c}=-(\kappa+i\Delta)\delta c-ig\cos^2(\theta/2)
c^{s}\delta
Q_{-}+\sqrt{2 \kappa}\delta c_{in},\vspace*{.1in}\\
\delta\dot{c}^{\dag}=-(\kappa-i\Delta)\delta
c^{\dag}+ig\cos^2(\theta/2) c^{s*}\delta Q_{-}+\sqrt{2 \kappa}\delta
c_{in}^{\dag}.
\end{array}
\end{equation}
Introducing the cavity field quadratures $\delta x=\delta c+\delta
c^{\dag}$ and $\delta y=i(\delta c^{\dag}-\delta c)$, and the input
noise quadratures $\delta x_{in}=\delta c_{in}+\delta c_{in}^{\dag}$
and $\delta y_{in}=i(\delta c_{in}^{\dag}-\delta c_{in})$, equation
(\ref{11}) can be rewritten in the matrix form
\begin{equation}\label{12}
\dot{f}(t)=Af(t)+\eta(t),
\end{equation}
in which $f(t)$ is the column vector of the fluctuations, $\eta(t)$
is the column vector of the noise sources. Their transposes are
\begin{equation}\label{13}
\begin{array}{lcl}
f(t)^{T}=(\delta Q_{-},\delta P_{-},\delta x,\delta y),\vspace*{.1in}\\
\eta(t)^{T}=(0,\xi_{1}-\xi_{2},\sqrt{2\kappa}\delta
x_{in},\sqrt{2\kappa}\delta y_{in});
\end{array}
\end{equation}
and the matrix $A$ is given by
\begin{equation}\label{14}
\fl A=\left(
  \begin{array}{cccc}
    0&\omega_{m} & 0 & 0 \\
    -\omega_{m}&-\gamma_{m} & -g\cos^2(\theta/2)(c^{s}+c^{s*})& ig\cos^2(\theta/2)(c^{s}-c^{s*}) \\
    -ig\cos^2(\theta/2)(c^{s}-c^{s*})  & 0 & -\kappa & \Delta \\
     -g\cos^2(\theta/2)(c^{s}+c^{s*})  & 0 & -\Delta & -\kappa \\
  \end{array}
\right).
\end{equation}
The solution of equation (\ref{12}) is $f(t)=M(t)f(0)+\int^{t}_{0}
M(t')\eta (t-t') dt'$, where $M(t)=e^{At}$. The system is stable and
reaches its steady state as $t\rightarrow \infty$ only if the real
parts of all the eigenvalues of the matrix $A$ are negative so that
$M(\infty)=0$. The stability conditions for the system can be found
by employing the Routh-Hurwitz criterion \cite{DeJesus}, we get
\begin{equation}\label{15}
\begin{array}{lcl}
\kappa\gamma_{m}[(\kappa^2+\Delta^2)^2+(2\kappa\gamma_{m}+\gamma_{m}^2-2\omega_{m}^2)(\kappa^2+\Delta^2)+\omega_{m}^2(4\kappa^2+\omega_{m}^2\vspace{.1in}\\
\hspace{.25in}+2\kappa\gamma_{m})]+2\omega_{m}\Delta g^2\cos^4(\theta/2)|c^{s}|^2(2\kappa+\gamma_{m})^2>0,\vspace{.1in}\\
\omega_{m}(\kappa^{2}+\Delta^2)-4\Delta
g^2\cos^4(\theta/2)|c^{s}|^2>0.
\end{array}
\end{equation}
All the parameters chosen in this paper have been verified to
satisfy the stability conditions (\ref{15}).

Fourier transforming each operator in equation (\ref{11}) by
$f(t)=\frac{1}{2\pi}\int^{+\infty}_{-\infty} f(\omega) e^{-i\omega
t} d\omega$ and solving it in the frequency domain, the relative
momentum fluctuations of the movable mirrors are given by
\begin{equation}\label{16}
\fl \eqalign{\delta
P_{-}(\omega)=\frac{i\omega}{d(\omega)}(2\sqrt{2\kappa}g\cos^2(\theta/2)\{[\kappa-i(\Delta+\omega)]c^{s*}\delta
c_{in}(\omega)+[\kappa+i(\Delta-\omega)]\vspace{.1in}\\
\hspace{.7in}\times c^{s}\delta
c_{in}^{\dag}(-\omega)\}-[(\kappa-i\omega)^2+\Delta^2][\xi_{1}(\omega)-\xi_{2}(\omega)]),}
\end{equation}
where $d(\omega)=-4\omega_{m}\Delta
g^2|c^{s}|^2\cos^4(\theta/2)+(\omega_m^2-\omega^2-i\gamma_{m}\omega
)[(\kappa-i\omega)^2+\Delta^2]$. Equation (\ref{16}) shows $\delta
P_{-}(\omega)$ has two contributions. The first term proportional to
$g$ originates from their interaction with the cavity field, while
the second term involving $\xi_{1}(\omega)$ and $\xi_{2}(\omega)$ is
from their interaction with their own environment. So the relative
momentum fluctuations of the movable mirrors are now determined by
radiation pressure and the thermal noise. In the case of no coupling
with the cavity field ($g=0$), the movable mirrors will make
Brownian motion only, $\delta
P_{-}(\omega)=-i\omega[\xi_{1}(\omega)-\xi_{2}(\omega)]/(\omega_{m}^2-\omega^2-i\gamma_{m}\omega)$,
whose mechanical susceptibility
$\chi(\omega)=1/(\omega_{m}^2-\omega^2-i\gamma_{m}\omega)$ has a
Lorentzian shape centered at the frequency $\omega_{m}$ with
$2\gamma_{m}$ as full width at half maximum (FWHM).

The mean square fluctuations in the relative momentum of the movable
mirrors are determined by
\begin{equation}\label{17}
\langle\delta
P_{-}(t)^{2}\rangle=\frac{1}{4\pi^{2}}\int\int_{-\infty}^{+\infty}
d\omega d\Omega e^{-i(\omega+\Omega)t} \langle\delta
P_{-}(\omega)\delta P_{-}(\Omega)\rangle.
\end{equation}

To calculate the mean square fluctuations, we require the
correlation functions of the noise sources in the frequency domain.
Fourier transforming equations (\ref{4}) and (\ref{5}) gives the
frequency domain correlation functions
\begin{equation}\label{18}
\begin{array}{lcl}
\langle\delta c_{in}^{\dag}(-\omega)\delta
c_{in}(\Omega)\rangle=2\pi N\delta(\omega+\Omega),\vspace{.1in}\\
\langle\delta c_{in}(\omega)\delta
c_{in}^{\dag}(-\Omega)\rangle=2\pi(N+1)\delta(\omega+\Omega),\vspace{.1in}\\
\langle\delta c_{in}(\omega)\delta
c_{in}(\Omega)\rangle=2\pi M\delta(\omega+\Omega-2\omega_{m}),\vspace{.1in}\\
\langle\delta c_{in}^{\dag}(-\omega)\delta
c_{in}^{\dag}(-\Omega)\rangle=2\pi M^{\ast}\delta(\omega+\Omega+2\omega_{m}),\vspace{.1in}\\
\langle\xi_{j}(\omega)\xi_{k}(\Omega)\rangle=2\pi\delta_{jk}\frac{\gamma_{m}
}{\omega_{m}}\omega\left[1+\coth(\frac{\hbar\omega}{2k_B
T})\right]\delta(\omega+\Omega).
\end{array}
\end{equation}
Upon substituting equation (\ref{16}) into  equation (\ref{17}) and
taking into account equation (\ref{18}), the mean square
fluctuations of equation (\ref{17}) are written as
\begin{equation}\label{19}
\fl \langle\delta
P_{-}(t)^{2}\rangle=\frac{1}{2\pi}\int_{-\infty}^{+\infty} [\omega^2
A+\omega (\omega-2\omega_m) B e^{-2i\omega_{m}t}+\omega
(\omega+2\omega_m) C e^{2i\omega_{m}t}] d\omega.
\end{equation}
where \begin{equation}\label{20}
\begin{array}{lcl}A=\frac{1}{d(\omega)d(-\omega)}(8\kappa
g^2\cos^4(\theta/2)|c^{s}|^2\{(N+1)[\kappa^2+(\Delta+\omega)^2]
\vspace{.1in}\\\hspace{.35in}+N[\kappa^2+(\Delta-\omega)^2]\}+2\gamma_{m}\frac{\omega}{\omega_{m}}[(\Delta^2+\kappa^2-\omega^2)^2+4\kappa^2\omega^2]\vspace{.1in}\\\hspace{.35in}\times[1+\coth(\frac{\hbar\omega}{2k_B
T})]), \vspace{.1in}\\
B=\frac{8\kappa g^2\cos^4(\theta/2)
c^{s*2}M}{d(\omega)d(2\omega_{m}-\omega)}
[\kappa-i(\Delta+\omega)][\kappa-i(\Delta+2\omega_{m}-\omega)],\vspace{.1in}\\
C=\frac{8\kappa g^2\cos^4(\theta/2)
c^{s2}M^{*}}{d(\omega)d(-2\omega_{m}-\omega)}[\kappa+i(\Delta-\omega)][\kappa+i(\Delta+2\omega_{m}+\omega)].\vspace{.1in}\\
\end{array}
\end{equation}
In equations (\ref{19}) and (\ref{20}), the term independent of $g$
is the thermal noise contribution; while all other terms involving
$g$ are the radiation pressure contribution, including the influence
of the squeezed vacuum light. Moreover, $\langle\delta
P_{-}(t)^{2}\rangle$ is time-dependent, the explicit time dependence
in equation (\ref{19}) can be eliminated by working in the
interaction picture. If we look the relative motion of the movable
mirrors as a harmonic oscillator and introduce the annihilation
(creation) operators $b$ ($b^\dag$) and $\tilde{b}$
($\tilde{b}^\dag$) for the oscillator in the Schr\"{o}dinger and
interaction picture with $[b,b^\dag]=1$ and
$[\tilde{b},\tilde{b}^\dag]=1$. They are related by $b=\tilde{b}
e^{-i\omega_{m}t}$ and $b^\dag=\tilde{b}^\dag e^{i\omega_{m}t}$.
Then using $P_{-}=i(b^\dag-b)$,
 and
$\tilde{P}_{-}=i(\tilde{b}^\dag-\tilde{b})$, we get
\begin{equation}\label{21}
\langle\delta
\tilde{P}_{-}^{2}\rangle=\frac{1}{2\pi}\int_{-\infty}^{+\infty}
[\omega^2 A+\omega (\omega-2\omega_m) B+\omega (\omega+2\omega_m) C]
d\omega.
\end{equation}

 According to equation (\ref{2}), the movable mirrors are said to be entangled if $\langle\delta{Q}_{+}^2\rangle$ and
 $\langle\delta \tilde{P}_{-}^2\rangle$ satisfy the inequality
\begin{equation}\label{22}
\langle\delta{Q}_{+}^{2}\rangle\langle\delta
\tilde{P}_{-}^{2}\rangle<1.
\end{equation}
where $Q_{+}=Q_{1}+Q_{2}$, the total displacement of the two movable
mirrors, which is  not related to the radiation pressure, only
determined by the thermal noise. At the temperature $T$, the
fluctuations $\langle\delta{Q}_{+}^{2}\rangle$ are
\begin{equation}\label{23}
\langle\delta{Q}_{+}^{2}\rangle=0.5+\frac{1}{e^{\hbar \omega_m/(k_B
T)}-1}
\end{equation}
Since $[Q_{+},P_{-}]=[Q_{1}+Q_{2},P_{1}-P_{2}]=0$, $Q_{+}$ and
$P_{-}$ can be simultaneously measured with infinite precision. Thus
$Q_{+}$ and $\tilde{P}_{-}$ can also be simultaneously measured with
infinite precision.

From equations (\ref{20}) and (\ref{21}), we find $\langle\delta
\tilde{P}_{-}^{2}\rangle$ is affected by the detuning $\Delta$, the
squeezing parameter $r$, the laser power $\wp$, the cavity length
$L$, the temperature of the environment $T$, and so on. In the
following, we confine ourselves to discussing the dependence of
$\langle\delta \tilde{P}_{-}^{2}\rangle$ on the squeezing parameter,
the laser power, and the temperature of the environment.

\section{Entanglement of the two movable mirrors}
In the section, we would like to numerically evaluate the mean
square fluctuations in the total displacement and the relative
momentum of the movable mirrors given by equations (\ref{23}) and
(\ref{21}) to show the entanglement of the two movable mirrors
produced by feeding the squeezed vacuum light at the input mirror.
To have fairly good idea of entanglement, we use the parameters of
a recent experiment \cite{Aspelmeyer} although we are aware that
the cavity geometry is different: the
wavelength of the laser $\lambda=\frac{2\pi c}{\omega_L}=1064$ nm,
$L=25$ mm, $m=145$ ng, $\kappa=2\pi\times215\times10^3$ Hz,
$\omega_m=2\pi\times947\times10^3$ Hz, the mechanical quality factor
$Q^{\prime}=\frac{\omega_{m}}{\gamma_{m}}=6700$, $\theta=\pi/3$.

\begin{figure}[!h]
\begin{center}
 \scalebox{0.65}{\includegraphics{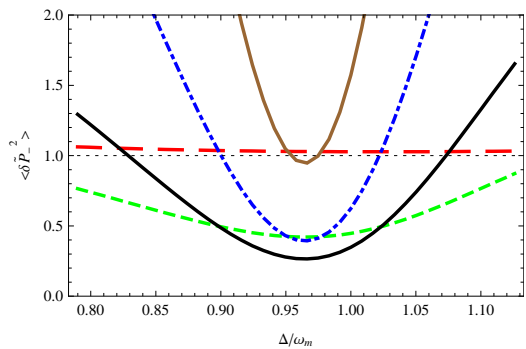}}
 \caption{\label{Fig2} The mean square fluctuations $\langle\delta \tilde{P}_{-}^{2}\rangle$
 versus the detuning $\Delta/\omega_{m}$ for different values of the squeezing of the input field.
  $r=0$ (red, big dashed line), $r=0.5$ (green, small dashed line), $r=1$ (black, solid curve),
   $r=1.5$ (blue, dotdashed curve), $r=2$ (brown, solid curve). The minimum values of
   $\langle\delta \tilde{P}_{-}^{2}\rangle$ are 1.027 ($r$=0), 0.420 ($r$=0.5), 0.265($r$=1),
0.394 ($r$=1.5), 0.947 ($r$=2). The flat dotted line represents
$\langle\delta \tilde{P}_{-}^{2}\rangle$=1. Parameters: the
temperature of the environment $T=41.4$ $\mu$K, the laser power
$\wp=3.8$ mW.}
\end{center}
\end{figure}

First we illustrate the squeezed vacuum light's effect on the
entanglement between the movable mirrors. We find as $T=41.4$
$\mu$K, the mean square fluctuations
$\langle\delta{Q}_{+}^{2}\rangle\approx1$, which implies that as
long as the mean square fluctuations $\langle\delta
\tilde{P}_{-}^{2}\rangle<1$, there is an entanglement between the
movable mirrors. The behavior of $\langle\delta
\tilde{P}_{-}^{2}\rangle$ at $\wp=3.8$ mW is plotted as a function
of the detuning $\Delta$ in figure ~\ref{Fig2}. Different graphs
correspond to different values of the squeezing of the input light.
In the case of no injection of the squeezed vacuum light ($r=0$),
which means that the squeezed vacuum light is replaced by an
ordinary vacuum light, we find $\langle\delta
\tilde{P}_{-}^{2}\rangle$ is always larger than unity, the minimum
value of $\langle\delta \tilde{P}_{-}^{2}\rangle$ is 1.027,
obviously there is no entanglement between the movable mirrors.
However, if we inject the squeezed vacuum light, it is seen that
entanglement between the movable mirrors occurs, meaning that there
is a quantum correlation between the movable mirrors, even through
they are separated in space. We also find the movable mirrors are
maximally entangled as the squeezing parameter is about $r=1$, the
corresponding minimum value of $\langle\delta
\tilde{P}_{-}^{2}\rangle$ is 0.265. So the injection of the squeezed
vacuum light leads to a significant reduction of the fluctuations in
the relative momentum between the movable mirrors. This is due to
the fact that using the squeezed vacuum light increases the photon
number in the cavity, which leads to a stronger radiation pressure
acting on the movable mirrors and enhances the entanglement between
the movable mirrors.

\begin{figure}[!h]
\begin{center}
\scalebox{0.65}{\includegraphics{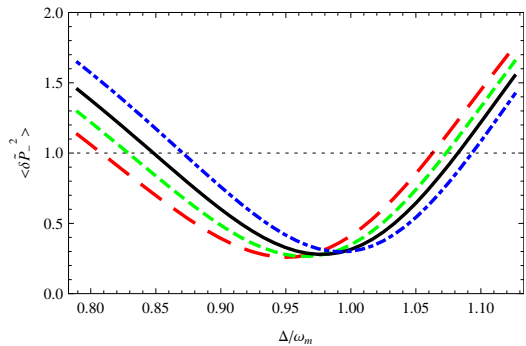}}
\caption{\label{Fig3} The mean square fluctuations $\langle\delta
\tilde{P}_{-}^{2}\rangle$ versus the detuning $\Delta/\omega_{m}$,
each curve corresponds to a different laser power. $\wp$=0.6 mW
(red, big dashed curve), 3.8 mW (green, small dashed curve), 6.9 mW
(black, solid curve), 10.7 mW (blue, dotdashed curve). The minimum
values of $\langle\delta \tilde{P}_{-}^{2}\rangle$ are 0.259
($\wp$=0.6 mW), 0.265 ($\wp$=3.8 mW), 0.279 ($\wp$=6.9 mW), 0.297
($\wp$=10.7 mW). The flat dotted line represents $\langle\delta
\tilde{P}^{2}_{-}\rangle$=1. Parameters: the squeezing parameter
$r=1$, the temperature of the environment $T=41.4$ $\mu$K.}
\end{center}
\end{figure}

Next we consider the influence of the laser power on the maximum
entanglement between the movable mirrors. We fix the squeezing
parameter $r=1$, and the temperature of the environment $T=41.4$
$\mu$K. We have already known at this temperature,
$\langle\delta{Q}_{+}^{2}\rangle\approx1$. Thus, if the mean square
fluctuations $\langle\delta \tilde{P}_{-}^{2}\rangle<1$, the movable
mirrors become entangled. The mean square fluctuations
$\langle\delta \tilde{P}_{-}^{2}\rangle$ as a function of the
detuning $\Delta$ for different laser power are shown in figure
~\ref{Fig3}. We find that significant entanglement occurs for a
range of pumping powers.
\begin{figure}[!h]
\begin{center}
\scalebox{0.65}{\includegraphics{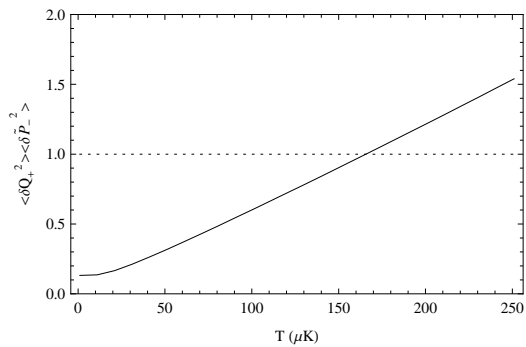}}
\caption{\label{Fig4} The value of
$\langle\delta{Q}_{+}^{2}\rangle\langle\delta
\tilde{P}_{-}^{2}\rangle$ versus the temperature of the environment
$T$ ($\mu$K). The minimum value of
$\langle\delta{Q}_{+}^{2}\rangle\langle\delta
\tilde{P}_{-}^{2}\rangle$ is 0.132 at $T=0$ K. The flat dotted line
represents $\langle\delta{Q}_{+}^{2}\rangle\langle\delta
\tilde{P}_{-}^{2}\rangle$=1. Parameters: the squeezing parameter
$r=1$, the laser power $\wp=3.8$ mW, the detuning $\Delta=0.965
\omega_{m}$.}
\end{center}
\end{figure}

We now show the effect of the temperature of the environment on the
entanglement between the movable mirrors. We fix the squeezing
parameter $r=1$, the laser power $\wp=3.8$ mW, and the detuning
$\Delta=0.965 \omega_{m}$. The value of $\langle\delta
Q_{+}^{2}\rangle\langle\delta \tilde{P}_{-}^{2}\rangle$ as a
function of the temperature of the environment is presented in
figure ~\ref{Fig4}. As the temperature of the environment increases,
the amount of entanglement monotonically decreases due to the
thermal fluctuations. This is as expected. What is remarkable is
that we find entanglement over a wide range of temperatures. As
$T\geq166$ $\mu$K, $\langle\delta Q_{+}^{2}\rangle\langle\delta
\tilde{P}_{-}^{2}\rangle\geq1$, the entanglement vanishes, the
movable mirrors become completely separable. So decreasing the
temperature of the environment can make the entanglement between the
movable mirrors stronger. Note that substantial progress has been
made in cooling the nanomechanical oscillators
\cite{Metzger,Naik,Gigan,Arcizet,Bouwmeester,Schliesser1,Thompson,Poggio,Corbitt,Groblacher,Schliesser}.
Further the ground state cooling using the resolved sideband regime
might soon become feasible. Clearly the entanglement depends on both
the quality factor of the cavity and the temperature of the
environment. The optical ring cavities are expected to yield much
higher quality factor: $\kappa\approx2\pi\times10$kHz, see for
example ~\cite{Klinner}, though for fixed mirrors replaced by moving
mirrors, the quality factor may be deteriorated. Metheods for
detection of entanglement are discussed in ~\cite{Mancini,Pinard}.
We note here that in our case we can deduce entanglement from the
knowledge of $\langle\delta \tilde{P}_{-}^{2}\rangle$. It can be
shown from equation (\ref{11}) that $\langle\delta
\tilde{P}_{-}^{2}\rangle$ can be obtained from the measurement of
the fluctuations in the quadrature of the output field.
\begin{figure}[!h]
\begin{center}
\scalebox{0.8}{\includegraphics{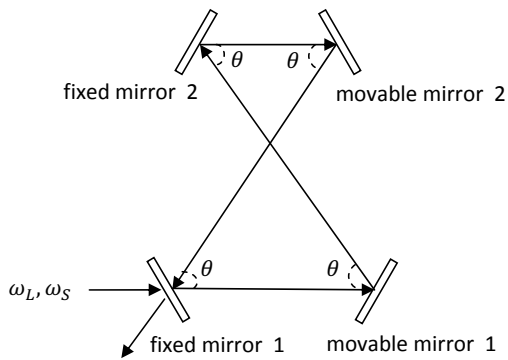}} \caption{\label{Fig5} Sketch
of 4-mirror ring cavity. A laser with frequency $\omega_{L}$ and
squeezed vacuum light with frequency
$\omega_{S}=\omega_{L}+\omega_{m}$ enter the ring cavity through the
partially transmitting fixed mirror 1. The fixed mirror 2 and the
two identical movable mirrors are perfectly reflecting.}
\end{center}
\end{figure}

If we use a different geometry of the ring cavity, as shown in
figure ~\ref{Fig5}, then we have the possibility of entangling other quadratures of the mirrors. In this case, the Hamiltonian of the system in the frame
rotating at the laser frequency becomes
\begin{equation}\label{24}
\begin{array}{lcl}
\eqalign{H=\hbar(\omega_{c}-\omega_{L})n_{c}-\hbar g
n_{c}\cos^2(\theta/2)
(Q_{1}+Q_{2})+\frac{\hbar\omega_{m}}{2}(Q_{1}^2+P_{1}^2)\\\hspace*{0.35in}+\frac{\hbar\omega_{m}}{2}(Q_{2}^2+P_{2}^2)+i\hbar\varepsilon(c^{\dag}-c),}
\end{array}
\end{equation}
We note the interaction between the two movable mirrors and the
cavity field depends only on the total displacement of the movable
mirrors. The movable mirrors are said to be entangled if
$\delta{Q}_{-}^2$ and
 $\delta \tilde{P}_{+}^2$ satisfy the inequality \cite{Mancini,Duan}
\begin{equation}\label{25}
\langle\delta{Q}_{-}^{2}\rangle\langle\delta
\tilde{P}_{+}^{2}\rangle<1.
\end{equation}
where $Q_{-}=Q_{1}-Q_{2}$ and $P_{+}=P_{1}+P_{2}$. The $Q_{-}$ is
the relative displacement of the two movable mirrors, which is not
related to the radiation pressure, only determined by the thermal
noise. The $P_{+}$ is the total momentum of the two movable mirrors,
and depends on the radiation pressure and the thermal noise. The
relation between $P_{+}$ and $\tilde{P}_{+}$ is the same as the
relation between $P_{-}$ and $\tilde{P}_{-}$ we defined above. Since
$[Q_{-},P_{+}]=[Q_{1}-Q_{2},P_{1}+P_{2}]=0$, $Q_{-}$ and $P_{+}$ can
be simultaneously measured with infinite precision. Thus $Q_{-}$ and
$\tilde{P}_{+}$ can also be simultaneously measured with infinite
precision. Through calculations, we find that
$\langle\delta{Q}_{-}^{2}\rangle$ and  $\langle\delta
\tilde{P}_{+}^{2}\rangle$ in a 4-mirror ring cavity have the same
form as $\langle\delta{Q}_{+}^{2}\rangle$ (equation (\ref{23}))
and$\langle\delta \tilde{P}_{-}^{2}\rangle$ (equation (\ref{21}))
 in a 3-mirror ring cavity, respectively. If we choose
the same parameters, the same numerical results will be obtained.
Therefore, using a 4-mirror ring cavity, the entanglement between
two oscillators can also be obtained.

\section{Conclusions}
In conclusion, we have found that the injection of squeezed vacuum
light and a laser can entangle the two identical movable mirrors
by the radiation pressure. The result shows the maximum
entanglement of the movable mirrors happens if the squeezed vacuum
light with $r$ about 1 is injected into the cavity. We also find
significant entanglement over a very wide range of input laser power
and temperatures of the environment.
\section *{Acknowledgement}
We gratefully acknowledge support for NSF Grants CCF 0829860 and
Phys. 0653494.

\end{document}